# Compatibility conditions, modulation mechanisms and preferred modes in incompressible flow over a cavity

N. Delprat[1,2]

*1 LIMSI-CNRS, BP133, 91403 Orsay Cedex, France*

*2 Université Pierre et Marie Curie, 75262 Paris Cedex, France*

ABSTRACT

Self-sustained oscillations in cavity-flows can be strongly influenced by shear layer instability acting together with feedback and modulation mechanisms. When coherently organized, these oscillations lock-on at a fundamental frequency and compatibility conditions exist between shear layer forcing, non linear interactions and low-frequency modulations. Special attention is given to the frequency coincidence which may appear in spectral distributions due to combinations between the dominant peak and its sidebands. Hence, the possible existence of two preferred modes in incompressible cavity-flows at medium Reynolds numbers is shown. This leads to a detailed categorization of the flow modulated regimes and to the specification of a persistent mode involved in modulation process whatever the oscillation stage.

*Introduction* - Besides the predominant frequency (denoted $f_a$ in the following), multiple peaks are generally present in spectra of the velocity fluctuations recorded in cavity-flows. They are usually referred to as secondary peaks even though their amplitude may be of the same order as the one of the primary frequency. Thus, sideband peaks around $f_a$ ($f_-, f_+$) and its harmonics $nf_a$ ($f_{-n}, f_{+n}$) have been identified as the result of an amplitude modulation of the fundamental oscillations by a low-frequency component (denoted $f_b$). Rockwell and Knisely proposed to relate this modulation process to variations in the vortex-corner impingement, assuming that its upstream influence is compatible with the whole oscillation system and feedback forcing of the shear layer instability [1]. Non-linear interactions between primary oscillation modes and secondary ones may also occur, yielding the generation of spectral components in the form of sum and difference frequencies and their combination. From a modulation analysis point of view, several interaction mechanisms which may coexist or combine with a feedback process must satisfy definite compatibility conditions and frequency selection rules for the oscillations to be self-sustained and coherent [2][3][4]. We propose to point out these conditions and to show how they may help to clarify the complex relationships between the frequencies involved in cavity-flows.

*Feedback cycle forcing and modulation mechanism* - As demonstrated by Miksad and co-authors for non-impinging shear layer excited by sound, non linear interactions between close instability modes $f_i$ and $f_{i'}$ generate amplitude and phase modulations through the generation of a low-frequency difference mode $d = (f_i - f_{i'})$ [5]. Resulting spectra have a multiple sideband distribution around the frequency of the most unstable mode $f_i$, predicted by linear stability theory, and its harmonics ($nf_i \pm d$, $n$ integer). Depending on the ratio of the interacting frequencies, some peaks may be superimposed and the number of spectral components is significantly reduced. For instance, when $f_{i'}/f_i = 3/4$, the second and fourth harmonics of $f_{i'}$ verify $2f_{i'} = 3/2f_i$, $4f_{i'} = 3f_i$ and the subharmonic of $f_i$ is such that $1/2f_i = 2/3f_{i'}$ [5]. For free shear layers artificially excited by a single frequency, spectral energy is partitionned between $f_i$, $nf_i$ and $1/2f_i$, $3/2f_i$ [6]. Note that the existence of spectral components at $1/2f_i$ and $3/2f_i$ is a specific feature of natural excited instability [7].

In cavity-flows, oscillations lock on a frequency selected in the band of dominant instability modes through a feedback mechanism between the two cavity corners. This frequency always lies in the neighbourhood of the frequency $f_i$ at which the shear layer is the most unstable [1]. As established from various experimental data, its corresponding Strouhal number based on the cavity length $L$ ($St_L = f_a L/U$ with $U$ free-stream velocity) is generally about $St_L \approx 1$ [8]. When expressed in terms of a locking condition (denoted $lc$), this means that the second oscillation mode of the phase-locked feedback cycle $f_{lc}$ is preferentially chosen by the cavity-flow [9]. As usually verified on the basis of phase measurements, this result requires that $U_c/U \approx 0.5$ and implies that $f_{lc} \approx U_c/L \approx U/2L$ (with $U_c$ convective velocity) [1][10]. Hence, the fundamental frequency of the cavity-flow oscillations satisfies $f_a \approx 2f_{lc}$. Note that in some experiments, cavity-flows have been found to oscillate in mode III, that is to say $St_L \approx 1.5$ and $f_a \approx 3f_{lc}$ [9].

When the predominant peak at frequency $f_a$ is modulated by a low-frequency $f_b$ such as $f_b/f_a \approx 0.4$ (stage II in the paper by Rockwell and Knisely [1]), it is of interest to point out that its associated Strouhal numbers $St_{\theta_0}$ is found to be lower than the ones predicted with inviscid stability theory ($St_{\theta_0} \approx 0.017$) [11][12]. It is rather in the same range than the Strouhal numbers of natural instability frequencies that is to say from 0.012 to 0.015 [13]. This feature is not specific to Rockwell's experiments. We have found similar Strouhal numbers with our experimental data obtained from LDV measurements for an air flow past an open cavity at medium Reynolds range numbers (free-stream velocities from 0.99 m/s to 2.84 m/s) and different cavity aspect ratios ($R = 1.5, 1.75, 2$) with ($L = 0.075$ m, $0.0875$ m, $0.1$ m) [14][15]. This remark is illustrated in table 1 where Strouhal numbers of predominant peaks and sideband ones ($f_-, f_+$) are given for several configurations. Strouhal numbers estimates of the fundamental feedback cycle in mode II ($2f_{lc} = U/L$) are also listed.

| configuration | $R = 2$ $U = 1.71$ m/s | $R = 2$ $U = 1.90$ m/s | $R = 2$ $U = 2.18$ m/s | $R = 1.5$ $U = 1.77$ m/s | $R = 1.5$ $U = 2.10$ m/s | $R = 1.5$ $U = 2.84$ m/s |
|---|---|---|---|---|---|---|
| $f_a \theta_0/U$ | 0.0132 | 0.0136 | 0.0116 | 0.0148 | 0.0146 | 0.0182 |
| $2(f_{lc}\theta_0/U)$ | 0.0126 | 0.0129 | 0.0107 | 0.0152 | 0.0146 | 0.0128 |
| $f_+\theta_0/U$ | 0.0175 | 0.0184 | 0.0155 | 0.0197 | none | none |
| $f_-\theta_0/U$ | 0.0084 | 0.0091 | 0.0076 | 0.0096 | 0.0124 | 0.0132 |

**Table 1** Strouhal numbers associated with predominant peaks ($f_a$), sideband peaks ($f_-, f_+$) and second harmonic of the fundamental feedback cycle ($2f_{lc}$). Details on the estimation of the momentum thickness $\theta_0$ at the cavity leading edge are available in the paper by Basley et al. (2010).

Despite the uncertainty in the initial momentum thickness $\theta_0$ due to its experimental estimation, the data confirm that the self-excited oscillations are forced by the feedback cycle (agreement within 4% between the two first lines in Table 1). For the ($R = 1.5$, $U = 2.84$ m/s) configuration, the shear layer oscillates in mode III ($3f_{lc}\theta_0/U \approx 0.0192$) at a frequency close to its inherent most unstable mode ($St_{\theta_0} \approx 0.0182$). Moreover, the lower Strouhal numbers observed for the ($R = 2$, $U = 2.18$ m/s) configuration can be explained by the fact that the cavity-flow regime is near a frequency jump between two stages of oscillation (transition from mode II to mode III) [1][10]. One may also easily verify that the $St_{\theta_0}$ value associated with the right sideband $f_+$ ($f_+ = f_a + f_b$) is equal to 0.0178 on average and that the $St_{\theta_0}$ value associated with the left sideband $f_-$ ($f_- = f_a - f_b$) is such that $2f_- \approx f_+$ (agreement within 2%). From an amplitude modulation point of view, this equality simply means that the modulating to fundamental oscillation frequency ratio satisfies $f_b/f_a \approx 1/3$. Except for the last two configurations, which will be discussed in detail later, the corresponding frequency ratios are found to be in good agreement with this value (cf Table 2). Note that for the cavity-flows under consideration, spectral components have been identified without ambiguity (as an illustration, see figure 1).

| configuration | $R = 2$ $U=1.71$m/s | $R = 2$ $U=1.90$m/s | $R = 2$ $U=2.18$m/s | $R = 1.5$ $U=1.77$m/s | $R = 1.5$ $U=2.10$m/s | $R = 1.5$ $U=2.84$m/s |
|---|---|---|---|---|---|---|
| $f_b$ (Hz) | 6.4 | 7.4 | 8.9 | 8 | 3.3 | 15.2 |
| $f_a$ (Hz) | 17.9 | 20 | 23.7 | 22.9 | 27 | 53.9 |
| $f_b/f_a$ | 0.36 | 0.37 | 0.38 | 0.35 | 0.12 | 0.28 |

**Table 2** Modulating frequency and fundamental oscillation frequency for different cavity-flow configurations. Reported values have been estimated from spectral distributions.

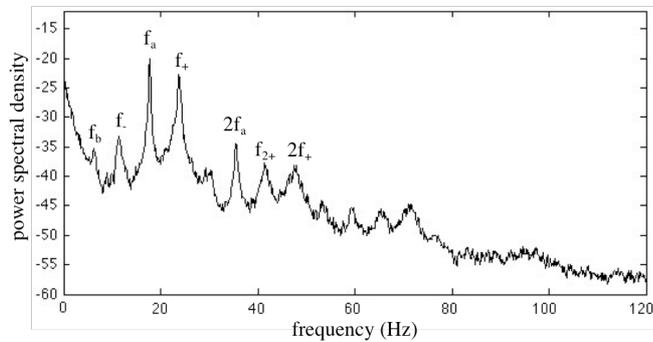

**Fig. 1** Spectral distribution obtained for the ($R=2$, $U=1.71$m/s) cavity-flow. Signals have been recorded near the downstream corner above the cavity top.

*Dual instability modes and frequency matching* - The presence of a low-frequency component $f_b$ equals to the difference frequency of the predominant peak and sidebands is typical of an amplitude modulation spectrum produced by non-linear interactions involving a low-frequency difference mode [5][16]. When this component satifies $f_b/f_a \approx 1/3$, it follows that $f_- \approx 1/2f_+$ and $f_a \approx 3/4f_+$. Hence, the frequency coincidence, previously described for a free shear layer artificially excited by two frequencies in the same ratio, is verified (see Table 3).

| configuration | $R = 2$ $U=1.71$m/s | $R = 2$ $U=1.90$m/s | $R = 2$ $U=2.18$m/s | $R = 1.5$ $U=1.77$m/s |
|---|---|---|---|---|
| $f_+$ (Hz) | 24 | 27 | 32.5 | 30.5 |
| $3/4f_+$ (Hz) | 18 | 20.25 | 24.37 | 22.88 |
| $f_a$ (Hz) | 17.9 | 20 | 23.7 | 22.9 |
| $2/3f_a$ (Hz) | 11.93 | 13.33 | 15.8 | 15.26 |
| $1/2f_+$ (Hz) | 12 | 13.5 | 16.25 | 15.25 |
| $2f_a$ (Hz) | 35.8 | 40.1 | 47.6 | 46.1 |
| $3/2f_+$ (Hz) | 36 | 40.5 | 48.75 | 45.75 |

**Table 3** Frequency matching between subharmonics of fundamental oscillation frequency $f_a$ and right sideband frequency $f_+$.

This remark suggests that spectral distributions may be described with the help of two frequency families which are: $(1/2f_+, f_+, 3/2f_+)$ and $(1/3f_a, 2/3f_a, f_a, 4/3f_a, 2f_a)$. In terms of non-linear interaction, this particular feature may be interpreted as the consequence of a dual-frequency excitation caused by two instability modes $(f_i, f_{i'})$. The first one would be due to the shear layer mode associated with the most amplified wave of the initial velocity profile ($f_+ \approx f_i$). The second one would be related to the instability mode whose frequency is compatible with the phase locking condition imposed by the cavity ($f_a \approx f_{i'}$). By analogy with the so-called preferred mode in jets [12][17], we propose to name the fundamental oscillation mode $f_a$ *the preferred mode of the cavity-flow* and the most unstable mode $f_i$ *the preferred mode of the shear layer*. We will see in the following that the differentiation between $f_i$ and $f_a$, even when they are very close in frequency, is essential to explain the connections between low-frequency modulations and flow lock-on features.

Our recent study on the modal spatial structures associated with the most characteristic frequencies of the ($R = 2$, $U = 1.90$m/s) cavity-flow has already shown that the contributions of $(f_a, f_+)$ and their harmonics are central to the global organization of the flow [15]. For the ($R = 2$, $U = 2.18$m/s) configuration, mode competition between $f_a$ and $f_+$ revealed by time-frequency and phase-space analyses of velocity signals also supports the presence of dual instability modes [18]. The existence of several admissible sources in mode generation expresses the well-known duality between non-linear interactions and wave modulations observed in free excited shear layers. In cavity-flow oscillations, the important point is that the excitation sources are inherent to the flow system and produce components which are compatible with the frequency selection rules of both excitation modes. This compatibility does not depend on the nature of their relationships (competition, combination or coexistence). As we will see in the next section, it is mainly related to the characteristics of the amplitude modulation process which may arise from their interaction.

*Amplitude-modulated regimes*   - A flow system which is self-excited by two instability modes may develop various dynamic behaviors. Depending on the frequency ratio and relative amplitudes of these modes, several regimes will be distinguished. In the present study, the frequency sequence ($1/3f_a$, $2/3f_a$, $f_a$, $4/3f_a$, $2f_a$) associated with the preferred mode $f_a$ of the cavity-flow is similar to the ones found for self-modulated impinging jets [19][20]. On the other hand, the frequency family related to $f_+ \approx f_i$ is typical of self-excited shear layer ($1/2f_i$, $f_i$, $3/2f_i$). Each family are representative of modulation processes which may combine through the non-linear interaction of $f_a$ and $f_i$ (see figure 2). Under this condition, several (modulated and non-modulated) regimes can governe the flow dynamics. Depending on the dual influence of the preferred modes, the flow may be mainly structured either by the free part of the shear layer, promoting non-linear interactions or by the oscillating system forced by the feedback mechanism, involving modulations [9]. In the boundary lock-on region, both kinds of regime may interact via the coupling of $f_i$ anf $f_a$ and a transitional state will appear in the form of mode switching. Whatever the case and even though they may physically compete in an intermittent way, the regimes will always spectrally act in harmony owing to their frequency coincidence.

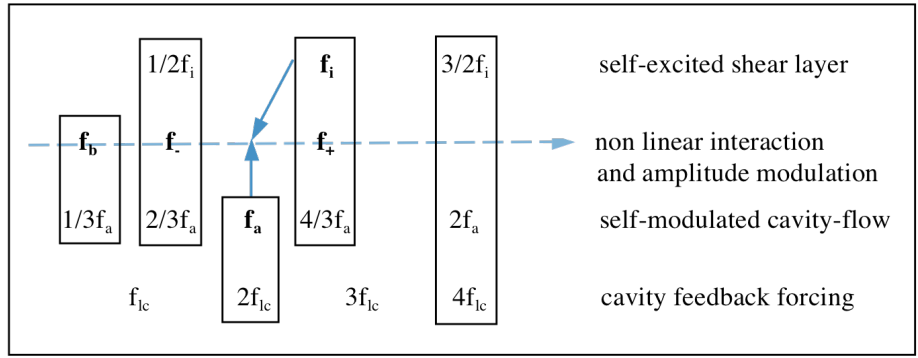

**Fig. 2** Schematic description of the possible combinations of the $f_a$ at $f_i$ frequency families when $f_a = 3/4f_i$ and $f_a = 2f_{lc}$. Frequency coincidence are indicated by rectangles. Note that $f_i$ is very close but not equals to $3f_{lc}$ ($3f_{lc} = 9/8f_i$).

The upstream influence of the amplitude fluctuations at the downstream corner strongly depends on the strength of these fluctuations. Most of the time, this influence is amplitude-modulated [1]. Since sideband energy comes from the modulating wave in amplitude modulation process, the $f_+$ mode may be reinforced through its interaction with the low-frequency mode $f_b$. Hence, the modulation process coud play a central role in the spectral transfer of energy between modes. This aspect is of particular interest in boundary regimes where frequency competition occurs between $f_a$ and $f_+$. It will be discussed more fully in a forthcoming paper with the help of time-frequency and bicoherence analyses.

*Very low-frequency modulations* - It is clear that the complexity and variety of modulated cavity-flow regimes cannot be exhaustively described by figure 2. For instance, there exist very low-frequency modulated regimes characterized by modulation ratios around 1/10. In this case, the sideband components around the fundamental oscillation frequency $f_a$ have to be differentiate from the ($f_-$, $f_+$) frequencies previously defined. For the sake of clarity, we propose to denote them $f_{-'}$ and $f_{+'}$ and to name the very-low frequency mode $\Delta_f$. The spectral distribution of the ($R = 1.5$, $U = 2.10$ m/s) cavity-flow depicted in figure 3 provides an illustration of this flow regime.

Here, the preferred mode of the cavity-flow ($f_a \approx 27$ Hz) and its harmonic are amplitude modulated by a low-frequency at $\Delta_f \approx 3.3$ Hz. Contrary to the $f_b$ modulation, the left sideband peak ($f_{-'} \approx 23.8$ Hz) has a greater amplitude than the right sideband ones ($f_{+'} \approx 30.2$ Hz). Barely discernable peaks at $f_{-'} \approx 16.1$ Hz and $f_{+'} \approx 38.9$ Hz might correspond to an extremely weak $f_b$ modulation process around 11.4 Hz (mean value). Note that if $f_- \approx 1/2 f_i$ (estimated $f_i \approx 32.53$ Hz), the $2f_- \approx f_+$ relation no longer holds. However, it is an interesting case because it reveals a second kind of modulation process whose effects might be masked by the $f_b$-modulation otherwise. Indeed, it has been established by one of us that two modulation processes present in an oscillating cavity-flow can spectrally combine when their modulating frequencies satisfy a specific compatibility condition (cf section 4 in [4]). This condition simply expresses the agreement between the $f_b$-modulation ratio $f_b/f_a$ and the modulating frequency ratio $\Delta_f/f_b$. In particular when $f_b/f_a = 1/3$, then $\Delta_f/f_a$ should be equal to 1/9 for the condition to be verified ($f_b/f_a = \Delta_f/f_b$). This value is strikingly close to the ones found in the present configuration ($\Delta_f/f_a \approx 0.12$).

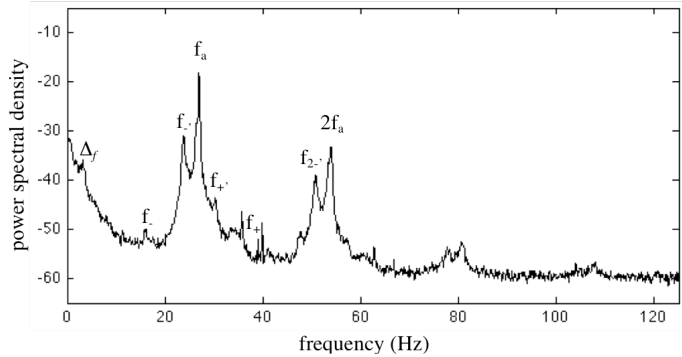

**Fig.3** Spectral distribution of the ($R = 1.5$, $U = 2.10$m/s) cavity-flow. Presence of a very low-frequency peak $\Delta_f$ and new sideband frequencies ($f_{-'}, f_{+'}$). Peaks associated with a weak $f_b$ modulation process ($f_-, f_+$) are barely discernable.

The result suggests that the $\Delta_f$-modulation could be on the same nature than the newly identified low-frequency modulation in compressible flows at subsonic Mach numbers and therefore be an inherent feature of cavity-flows [4][21]. If as supposed in the compressible case, the very low-frequency mode is linked to the internal flow then the studied spectral distribution could simply represent a recirculation induced modulation. Thus, the global dynamics of the flow could be structured by oscillations at the fundamental frequency $f_a$, modulated via delayed non-linear interactions through the recirculating flow. An extensive study of the local and global properties of the $\Delta_f$ and $f_{-'}$ modes has been recently performed and their relationships with the $f_a$ mode will be described in detail in a separate paper.

*Cavity-flow resonances and persistent mode* - When a cavity-flow oscillates in mode III, the change in phase locking condition ($f_a \approx 3f_{lc}$) leads to noticeable modifications in spectra with regard to the spectral features examined in section 2. The ($R = 1.5$, $U = 2.84$m/s) configuration listed in Table 1 illustrates this kind of flow regime with $f_a \approx 53.9$ Hz and $f_{lc} \approx 18.9$ Hz. Its spectral distribution shown in figure 4 exhibits a well-defined low-frequency peak at the difference frequency between $f_a$ and its left sideband peak $f_-$ (cf Table 4). Since the corresponding modulation ratio lies between 1/4 and 1/3 ($f_b/f_a \approx 0.282$), we have identified it as the component $f_b$. The preferred mode of the shear layer and its subharmonic are also present ($f_i \approx 50.3$ Hz from linear stability analysis). Contrary to regimes in mode II, the $f_i$ peak frequency is slightly smaller than the $f_a$ ones (about 4 Hz) and no right sideband peak at $f_+$ is detected.

At high free-stream velocities, increase of the random background fluctuations as well as strong non-linear activity are expected to occur. The observed spiky structure of the spectrum in figure 4 expresses the former feature. The latter is confirmed by the multiple sum and difference modes, which may contribute to the existence of some peaks. For instance, the following combinations $(f_- + f_a)$, $(2f_a - f_b)$ or $(2f_- + f_b)$ may generate the $f_7$ frequency. Another interesting characteristic is the extended influence of the cavity feedback forcing on the sideband component $f_-$, which is one of the predominant peaks in the spectrum. Indeed, $f_-$ is very close to the second harmonic of $f_{lc}$ ($2f_{lc} \approx 37.8$ Hz) and $2f_-$ to its forth ones ($4f_{lc} \approx 75.6$ Hz). Note also that $5f_{lc} \approx 94.5$ Hz ($f_7 \approx 92.2$ Hz). Thus, the cavity-flow resonances at ($2f_{lc}$, $3f_{lc}$, $4f_{lc}$, $5f_{lc}$) seem to strongly structure the flow behavior. In particular, the large amplitude of the $f_-$ mode is apparently due to its frequency coincidence with mode II of the fundamental feedback cycle.

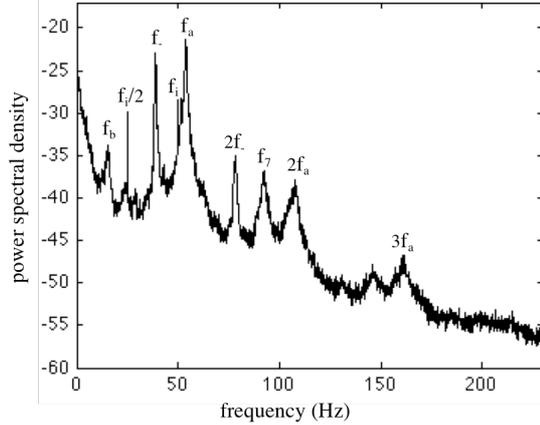

**Fig.4** Spectral distribution of the ($R = 1.5$, $U = 2.84$m/s) cavity-flow. Dominant mode $f_a$ corresponds to mode III of the feedback cycle $f_{lc}$.

| $f_b$ (Hz) | $1/2f_i$ (Hz) | $f_-$ (Hz) | $f_i$ (Hz) | $f_a$ (Hz) | $2f_-$ (Hz) | $f_7$ (Hz) | $2f_a$ (Hz) |
|---|---|---|---|---|---|---|---|
| 15.2 | 24.9 | 38.8 | 49.7 Hz | 53.9 | 78.2 | 92.2 | 107.7 |

**Table 4** Frequencies identified in the spectrum of the ($R = 1.5$, $U = 2.84$m/s) cavity-flow.

In boundary regimes oscillating in mode II, we have established that the dominant peaks are located at $f_a$ and $f_+$ with $f_a \approx 3/4f_i \approx 2f_{lc}$ and $f_+ \approx f_i$. In the present configuration, they are positionned at $f_- \approx 2f_{lc}$ and $f_a \approx 3f_{lc}$ with $f_- \approx 3/4f_i$ (see Table 5). This indicates that the peak at $3/4f_i$ is always involved in modulation process. When non-linear effects are sufficiently significant, it may interact with another dominant mode to generate an amplitude modulation at their difference frequency. The modulation mechanism will be triggered under two conditions. Firstly, the amplitudes of the interacting modes have to be large enough for the interaction to be efficient. This will be obviously verified when both modes lie in the vicinity of a cavity-flow resonance. Secondly, their difference frequency must be in tune with the modulating frequency inherent to the modulated impinging flow, that is to say $f_b$ must be in the neighborhood of $1/3f_a$. These remarks make it appropriate to think that the $3/4f_i$ mode could play a central role in the coupling between non-linear aspects of the shear layer and self-modulation of the cavity-flow. Its persistence between two stages of oscillation is an additional argument for considering its contribution as an essential feature of the oscillation mechanism. For that reason, it will be called *the persistent mode* and denoted $f_{pm}$.

| $2f_{lc}$ (Hz) | $3/4f_i$ | $4f_{lc}$ (Hz) | $3/2f_i$ |
|---|---|---|---|
| 37.8 | 37.5 | 75.6 | 75 |

**Table 5** Frequency coincidence between cavity-flow resonances and subharmonics of the $f_i$ mode.

***Concluding remarks*** - The above considerations allow us to clearly characterize each stage of oscillation through their modulation features. For stage II, the frequencies involved in the modulation process are such that:

$$f_a \approx f_{pm} \quad \text{and} \quad f_b \approx f_+ - f_a \qquad \text{with } f_+ \approx f_i \text{ and } f_b/f_a \approx 1/3. \qquad (1)$$

In terms of an amplitude modulation, this means that the sideband peaks ($f_{-n}, f_{+n}$) may simply be predicted by:

$$f_{\pm n} = (nf_a \pm f_b) \approx (n \pm 1/3) f_a \qquad n \text{ integer}, n \geq 1. \qquad (2)$$

In stage III, we have:

$$f_- \approx f_{pm} \quad \text{and} \quad f_b \approx f_a - f_-. \qquad (3)$$

Morover, the preferred mode frequencies $f_i$ and $f_a$ are very close. For instance, in the ($R = 1.5$, $U = 2.84$m/s) configuration, $f_i/f_a \approx 0.928$. From these results, it follows that the corresponding modulation ratio may be written in the form:

$$f_b/f_a \approx 1 - 3/4\, f_i/f_a. \qquad (4)$$

Eq. (4) allows us to quantify the observed deviation of the $f_b/f_a$ ratio from it expected value 1/3, while remaining close to it. It is clear that reasonably good estimates can also be found when the frequency difference between $f_a$ and $f_i$ is not taken into account. In this case, the modulation ratio is approximated by $f_b/f_a \approx 1/4$ and the sideband peaks are defined by:

$$f_{\pm n} \approx (n \pm 1/4) f_a \qquad n \text{ integer}, n \geq 1. \qquad (5)$$

Expressed in terms of a Strouhal number, Equation (5) becomes:

$$f_{\pm n} L/U \approx \kappa_3 (n \pm \gamma) \qquad \text{with } \gamma = 1/4 \text{ and } \kappa_3 = 3\kappa \ (\kappa \approx 0.5). \qquad (6)$$

It follows that the sequence of the left sideband peaks is predicted by:

$$f_{-n} L/U \approx \kappa_3 (n - \gamma) \qquad \text{with } \gamma = 1/4 \text{ and } \kappa_3 = 3\kappa = 1.5 \qquad (7)$$

Eq. (7) does not exactly correspond to the Rossiter formula [22] applied in the zero Mach number limit, which is:

$$f_{-n} L/U \approx \kappa (n - \gamma) \qquad \text{with } \gamma = 1/4 \text{ and } \kappa = 0.57. \qquad (8)$$

However, it is to be pointed out that when $n = 2$ in Eq. (8) the dimensionless frequency is close to 1 ($St_L \approx 0.99$). Thus, the associated mode can be identified as the $f_-$ mode, defined in Eq. (7) with $n = 1$ ($St_L \approx 1.12$). Note also that $n = 3$ in Eq. (8) will give the estimated Strouhal number associated with $f_a$ ($St_L \approx 1.56$). In state II, the same can be verified for $f_a$ ($St_L \approx 1$) and $f_+$ ($St_L \approx 1.33$).

This remark gives a key to explain why the usual extension of the Rossiter formula to incompressible cavity-flows may provide an empirical way of predicting dominant mode frequencies with standard values ($\gamma = 1/4$, $\kappa = 0.57$) and $n \geq 2$ (see for instance [23] and references therein). It also highly supports the view that modulation processes with similar characteristics could be involved in compressible and incompressible flows, even though the primary oscillation frequencies are not identically defined owing to the difference in the nature of the feedback mechanism (aeroacoustic loop or hydrodynamic forcing).